\documentstyle{article}
\newcommand{\lm}[1]{\mbox{$\lambda_{#1}$}}
\newcommand{\m}[1]{\mbox{$m_{#1}^{2}$}}

\renewcommand{\H}[1]{\mbox{$H_{#1}$}}
\newcommand{\Hd}[1]{\mbox{$H_{#1}^{\dag}$}}

\newcommand{\e}{\mbox{electroweak }}

\begin{document}
\begin{flushright}
GUTPA/97/11/3
\end{flushright}
\vskip .1in

\begin{center}

{\Large \bf Spontaneous CP violation and Higgs spectrum in a
Next to Minimal Supersymmetric Model. }

\vspace{50pt}

{\bf A.T.\,Davies, C.D.\,Froggatt and A.\,Usai}

\vspace{6pt}

{ \em Department of Physics and Astronomy\\
 Glasgow University, Glasgow G12 8QQ,
Scotland\\}
\end{center}

\section*{ }
\begin{center}
{\large\bf Abstract}
\end{center}

We explore the possibility of spontaneous CP violation within the
Next to Minimal Supersymmetric Standard Model.  In the most general
form of the model, without a discrete $Z_3$ symmetry, we find that
even at tree level  spontaneous CP violation can occur,  while also
permitting Higgs masses consistent with experiment.

\vskip 4.0cm
{\em To be published in the Proceedings of the
International Europhysics Conference on High Energy Physics,
Jerusalem, August 1997.}

\thispagestyle{empty}
\newpage

\section{NMSSM} \label{NMSSM}

CP violation, the Higgs spectrum and supersymmetry are at the
forefront of experimental investigation and theoretical interest.
We focus on spontaneous CP violation in the Next to Minimal
Supersymmetric Standard Model (NMSSM) which contains a singlet
$N$  in addition  to the two doublets  $H_1$ and $H_2$ of the
MSSM.  Spontaneous CP violation is achievable in both the MSSM
and NMSSM with the usual $Z_3$ discrete symmetry, but only by
invoking radiative corrections to raise a negative
$\mbox{(mass)}^2$ mode to a small, experimentally unacceptable,
real mass \cite{ Maekawa, Pomerol, Babu}

Our main result is that spontaneous CP violation is possible
for the general NMSSM potential {\em even at tree level},
and that Higgs masses need not be small.

Above some SUSY-breaking scale, $M_S$, unknown but hopefully not
far beyond experimental reach, the most general superpotential
for these fields is \cite{GunHHG,GunHab}
\begin {eqnarray}
W =  \lambda NH_1H_2 -\frac{k}{3} N^3 - r N + \mu H_1H_2 + W_{Fermion}
\end{eqnarray}
$\mu$ has dimension of  mass and $r$ of $\mbox{(mass)}^2$.
At lower energy  a general quartic form is adopted, with
8 couplings $\lambda_i$, which at $M_S$  may be expressed
in terms of the gauge couplings and the superpotential
coupling constants, and may be determined at the electroweak
scale $M_{Weak}$ using renormalization group (RG) equations,
if  $M_S  > M_{Weak}$.
New SUSY-breaking soft cubic and quadratic terms are added.
In  much of the work on the NMSSM,  RG equations are used
to run down  the soft couplings from a hypothesized universal
form at the scale  $M_{GUT}$, but we do not assume this,
and regard $ m_i, i = 1 \dots 7 $, below as arbitrary parameters.

The effective potential is  thus \cite{GunHHG,GunHab}
\begin{eqnarray}
\label{eq.vs0}
V_{0} & = & \frac{1}{2}\lm{1}(\Hd{1}\H{1})^2 +
\frac{1}{2}\lm{2}(\Hd{2}\H{2})^2   \nonumber  \\
  &  & + (\lm{3}+\lm{4})(\Hd{1}\H{1})(\Hd{2}\H{2}) -
\lm{4}\left| \Hd{1}\H{2}\right|^{2} \nonumber\\
  &  & + (\lm{5}\Hd{1}\H{1} +\lm{6}\Hd{2}\H{2})N^{\star}N+
(\lm{7}\H{1}\H{2}N^{\star 2}+h.c.)  \nonumber  \\
   &  &+  \lm{8}(N^{\star}N)^2+
\lambda \mu (N+h.c.)(\Hd{1}\H{1}+\Hd{2}\H{2})  \nonumber \\
 & & + \m{1}\Hd{1}\H{1}+\m{2}\Hd{2}\H{2} + \m{3}N^{\star}N -
m_4 (\H{1}\H{2}N+h.c.)
 \nonumber\\
 & & - \frac{1}{3}m_5(N^3+h.c.)+\frac{1}{2}
\m{6}(\H{1}\H{2}+h.c.)+ \m{7}(N^2+h.c.)
\end{eqnarray}

A restricted version of this has been advocated to explain
why  $\mu$ is of electroweak scale, the `$\mu$-problem' of
the MSSM.  The terms in the superpotential involving
dimensionful couplings and the soft terms in $m_6$ and
$m_7$ are dropped, leaving a $Z_3$ discrete symmetry which
protects the hierarchy. The VEV $<N>$ replaces $\mu$,
thereby introducing  a domain wall problem \cite{EllisGHRZ,AbelSW}.
This restricted model  provides only a limited improvement
on the MSSM:

(a) Surveys of the parameter space tend to favour a low energy
Higgs spectrum  very similar to the MSSM \cite {KingW, Ellwanger}

(b)  As in the MSSM, spontaneous CP violation is possible, but
only as a result of radiative
corrections \cite{Romao, Asatrian1, Haba}. The lightest
neutral Higgs has a mass less than  45 GeV, and
typically $\mbox{tan} \beta \le 1$.

We do not impose the $Z_3$ symmetry, so do not have the above
domain wall problem, but are left with the $\mu$-problem.
We find that spontaneous CP violation is more easily achieved
in this model than in the MSSM or the  NMSSM with $Z_3$.

We consider  real coupling constants, so that the tree level
potential is CP conserving, but admit complex  VEVs for the
neutral fields, giving
\begin{eqnarray} \label{eq.vsvac}
V_{0} &  = &  \frac{1}{2}(\lm{1}v_1^4  +\lm{2}v_2^4) +
(\lm{3}+\lm{4})v_1^2 v_2^2  +
(\lm{5}v_1^2 +\lm{6}v_2^2)v_3^2  \nonumber\\
&  & + 2\lm{7}v_1 v_2 v_3^2 cos(\theta_1 + \theta_2 -
2 \theta_3)+ \lm{8}v_3^4 +
2 \lambda\mu (v_1^2+v_2^2)v_3 cos(\theta_3)  \nonumber\\
 &  & + \m{1}v_1^2+\m{2}v_2^2 +
\m{3} v_3^2-2m_4 v_1 v_2 v_3 cos(\theta_1 + \theta_2 + \theta_3)
 \nonumber\\
 & & - \frac{2}{3} m_5 v_3^3 cos(3 \theta_3)+
2\m{6}v_1 v_2cos(\theta_1 + \theta_2) +2\m{7} v_3^2 cos(2 \theta_3)
\end{eqnarray}
\begin{eqnarray}
\langle H_i^0 \rangle = v_i e^{i \theta_i} (i=1,2),
\hspace{1cm} \langle N \rangle=v_3 e^{i \theta_3}.
\end{eqnarray}
where, without loss of generality,  $\theta_2 = 0$,
and $0 \le \theta_3 < \pi$.

This potential has 3 extra terms as compared with a $Z_3$
invariant potential: a cubic term arising from the $\mu$
and $\lambda$ terms in the superpotential, and two quadratic
terms with coefficients $m_6^2$ and $m_7^2$.  If $\mu = 0$
the effective potential  loses this cubic term, and the
quadratic terms alone account for the difference between our
results and previous ones in the literature.

The scalar mass matrix gives rise to 1 charged and 5  neutral
particles.  If  the angles $\theta_1 \mbox{ and } \theta_3 $
are non-zero,  the neutral matrix does not decouple into
sectors with CP = +1 and -1.

\section {Searches and results} \label{Results}

We search the parameter space of this potential  to see
if spontaneous CP violation is compatible with experimental
bounds on the Higgs spectrum.

There are 11 parameters, but we impose some restrictions.

 (a)  4 superpotential parameters  $\mu, \lambda, k,
\mbox{ and } h_t, \mbox{ the top Yukawa coupling} $:
\\Running $(\lambda, k)$ up  from the \e scale using RG
equations \cite{Ekw2} requires them to be small  to
avoid blow-up at high energy. In the examples below we
fix $\lambda=0.5, k = 0.5$.  We take the fixed point
value $h_t  = 1.05$ at the SUSY scale, bearing in mind that
at the \e scale the running top mass
is  $h_t v_0 \mbox{sin} \beta$.

(b)  Soft breaking mass parameters  $m_i, i=\dots7$: \\
Five of these can be traded for VEV magnitudes and phases
$v_1,v_2,v_3,\theta_1, \theta_3$.
We  eliminate one by the condition $v_0 =
\sqrt{v_1^2 + v_2^2} = 174$ GeV
and replace 2 others by  the conventional  \mbox{$\tan{\beta }
\equiv v_2/v_1$} and \mbox{$R \equiv v_3/v_0 $}.
A sixth mass, $m_4$ say, can be exchanged for the mass of the
charged Higgs, $M_{H^+}$, which
\mbox{ $b \rightarrow s + \gamma$} experiment suggests to be
greater than 250 GeV \cite{CLEO}.
We can obtain an analytic form for this:
\begin {eqnarray}
M_{H^+}^2 & = & - \lambda_4 v_0^2 -
\frac{2(\lambda_7 v_3^2sin (3 \theta_3) +
m_6^2 sin\theta_3)}{sin(2\beta) sin(\theta_1+\theta_2+\theta_3)}
\end{eqnarray}
This shows how the parameter $m_6^2$, not necessarily positive,
introduces extra freedom to raise the charged Higgs mass.
This leaves one parameter $m_5$, with no particular interpretation.

In the case $\mu=0 $, $\lambda$ and $k$ cannot be too small.
In the simplest case, where the Higgsino-gaugino mixing is
small, the (unbroken SUSY) charged Higgsino mass is
$|\mu+ \lambda v_3 \mbox{cos} \theta_3|$. The experimental
lower bound, conservatively
$ \frac{1}{2}M_Z$,  then disallows small $R\  (\equiv v_3/v_0)$.
There is no upper bound on $R$, and indeed the cubic and quartic
terms in the field N can in some cases provide deep global
minima of the potential for large $R$, thus excluding some
otherwise acceptable values of the other parameters.

Our {\em modus operandi} was to scan over a grid of  parameters
chosen to make the first derivatives of the potential vanish at
prescribed VEVs.  Numerical minimization was  performed to
ensure that these were minima, giving positive
$ \mbox{(mass)}^2$, and to reject local minima. In many
apparently reasonable cases  the spontaneous CP violating
minimum was metastable, with a lower electroweak and  CP
conserving vacuum lying at $v_1=v_2 = 0$ and $v_3$
the order of  TeV.  We present two indicative examples from
preliminary searches.

\vspace{0.25 cm}
\begin{tabular} {|c|c|c|c|c|c|c|c|} \hline
\multicolumn{8}{|c|} {\bf Parameters giving spontaneous
CP violation} \\
\hline
 CASE & \mbox{tan}$\beta$ & R   &
$\theta_1$  &  $\theta_3$  &  $M_{H^+}$  & $m_5$  &  $\mu$\\
\hline
A & 2.0 & 2.0  & 1.20$\pi$ & 0.65$\pi$  & 250 GeV &
60 GeV & -20 GeV\\
\hline
B & 2.0 & 2.0 & 0.60$\pi$ & 0.35$\pi$ & 250 GeV & -100 GeV & 0 \\
\hline
\end{tabular}
\vspace{0.25 cm}

Case (A) is for a  SUSY scale of  1 TeV, with quartic couplings
radiatively corrected using  RG equations, assuming that all
squarks and gauginos lie at the SUSY scale. This has neutral
Higgs massses from 89 to 318 GeV, a lightest neutralino of
48 GeV and a chargino of 99 GeV.  Larger neutralino masses
can be obtained by increasing R.

Case (B) is for $M_S= M_{Weak}$ i.e. a  SUSY quartic potential.
It gives  neutral Higgs masses of  81 to 372 GeV, a lightest
neutralino of 47 GeV, and a chargino of 79 GeV, ignoring gaugino
mixing which should be included if indeed the SUSY scale is as
low as the \e scale.

This case is presented as a go-theorem, a counter example to
no-go examples \cite{Romao}. As $\mu = 0$ here,  the quartic
potential has the standard $Z_3$ invariant form and avoids the
conclusion of Romao \cite{Romao} due to the two $Z_3$ violating
soft terms. Nor do the conditions of the Georgi-Pais \cite{Georgi}
no-go theorem apply. It relates to the situation where CP is
conserved at tree level and is broken only by perturbative
radiative corrections. The soft terms here are not in this category.

We are encouraged to explore further the phenomenology of such models.

% ---- Bibliography ----
%

\end{document}